\documentstyle[prb,aps,psfig]{revtex}
\begin{document}
\title{Reply to the Comment on: Quantum Monte Carlo study of the dipole moment of CO [J. Chem. Phys. 110, 11700 (1999)]. }
\author{F. Schautz and H.-J. Flad$^*$}
\maketitle
\begin{center}
Max-Planck-Institut f{\"u}r Physik komplexer Systeme, \\
N\"othnitzer Str. 38, D-01187 Dresden, Germany
\end{center}
\begin{flushleft}
$^{*}$ present address: Max-Planck-Institut f\"ur Mathematik in den Naturwissenschaften, \\
Inselstr. 22-26, D-04103 Leipzig, Germany \\
e-mail:flad@mis.mpg.de\\
FAX: ++49-(0)-341-9959-999
\end{flushleft}
\vspace{10mm}

\noindent 
In Ref. 1 we have mistakenly claimed that the 
applicability of the Hellmann-Feynman theorem in fixed-node quantum Monte Carlo calculations is not subject to the 
manner how the nodal boundary depends on an external parameter $\lambda$.
As it has been pointed out by Huang et al. \cite{X2}
in their comment on Ref. 1, this statement is not correct in general, except where the Hellmann-Feynman
force is calculated for a nodal boundary which coincides with that of the unconstrained exact
eigenfunction. We want to point out the error in our arguments and present an explicit expression 
for the correction term which supplements the Hellmann-Feynman force.

In our approach the fixed-node approximation is treated as a Dirichlet
type of boundary value problem \cite{X4} on a nodal region $\Omega$. 
Properly stated $\Omega$ has to be taken as an open subset of $R^{3N}$
in which the fixed-node wavefunction $\Psi(\lambda,\Omega)$ satisfies the Schr\"odinger equation 
\begin{equation}
\hat{H}(\lambda) \; \Psi(\lambda,\Omega) = E[\lambda,\Omega] \; \Psi(\lambda,\Omega) .
\label{eq7}
\end{equation}
The fixed-node wavefunction $\Psi(\lambda,\Omega)$ has to vanish on the 
boundary $\partial \Omega$ of $\Omega$ and must be continous on the entire $\Omega \, \cup \, \partial \Omega$.
In order to define the derivatives
of the wavefunction on the boundary $\partial \Omega$ one has to take the limit
of derivatives of interior points when approaching the boundary \cite{X3}. 
Within such a framework no $\delta$ function term appears for 
the second derivatives on $\partial \Omega$ since we do not extend our wavefunction beyond the boundary.
Neclecting spurious singularities in the potential $\hat{V}(\lambda)$ on $\partial \Omega$
we can actually conclude from 
\begin{equation}
\Delta \Psi(\lambda,\Omega) = 2 \left( \hat{V}(\lambda) - E[\lambda,\Omega] \right) \Psi(\lambda,\Omega)
\label{eq8}
\end{equation}
that the limits of the second derivatives vanish almost everywhere on the boundary.
We want to stress however that this is not in contradiction to the arguments given in Ref. 2,
where the wavefunction has been extended over the whole space. In this case one actually
encounters discontinous first derivatives when crossing the nodes.

Revisiting Eq. (8) of Ref. 1, we can identify the missing boundary term mentioned in Ref. 2 by inspection of the second   
line. After differentiation with respect to $\lambda$ we obtain
\begin{equation}
\int_{\Omega(0)} \left[ \partial_\lambda \Psi(0) \; \hat{H}_0 \; \Psi(0)  
                       +\Psi(0) \left. \partial_\lambda \hat{H}_\lambda \right|_{\lambda=0} \Psi(0)
                       +\Psi(0) \; \hat{H}_0 \; \partial_\lambda \Psi(0)  \right]  d\Omega 
\label{eq1}
\end{equation}
The first term in the integral vanishes due to Eq. (~\ref{eq7}) and the normalization constraint Eq. (6) in Ref. 1
from which follows \footnote{To see that this is valid for a parameter dependent boundary we refer to Ref. 1 for
a discussion of the boundary terms.}
\begin{equation}
\int_{\Omega(0)} \Psi(0) \; \partial_\lambda \Psi(0) \; d\Omega = 0 ,
\label{eq9}
\end{equation}
whereas the second term yields the standard Hellmann-Feynman force, however the third 
term does not vanish in general as we have erroneously claimed in Ref. 1. Applying Green's
second formula \cite{X4} we can rewrite the third term 
\begin{equation}
\int_{\Omega(0)} \Psi(0) \; \hat{H}_0 \; \partial_\lambda \Psi(0) \; d\Omega 
= -\frac{1}{2} \int_{\partial \Omega(0)} \mid \! \nabla \Psi(0) \! \mid \partial_\lambda \Psi(0) \; d \partial \Omega
\label{eq2}
\end{equation}
where the sign of the boundary term corresponds to $\Psi > 0$ inside the nodal region.
For this step we presume that the first and second derivatives of $\Psi(0)$ and $\partial_\lambda \Psi(0)$ 
can be continuously extended to the boundary in the sense discussed above. 
In order to get a better understanding of the physical character of this term we have to generalize our 
considerations by allowing the external parameter $\lambda$ and the nodal domain $\Omega$ to vary 
independently. Doing so we can rewrite the derivative $\partial_\lambda \Psi(0)$ like a total differential
\begin{equation}
\label{eq3}
\partial_\lambda \Psi(0) =  
 \left. \left[ \; \partial_\lambda \Psi(\lambda,\Omega(0)) 
+  \partial_\lambda \Psi(0,\Omega(\lambda)) \; \right] \right|_{\lambda=0} 
\end{equation}
The first term $\partial_\lambda \Psi(\lambda,\Omega(0))$ corresponds to the change of the wave function 
with respect to the external parameter $\lambda$ under the constraint that the nodes are kept fixed.
It vanishes on the boundary and therefore does not contribute to Eq. (~\ref{eq2}).
The second term represents the change of the wave function $\partial_\lambda \Psi(0,\Omega(\lambda))$ 
under a variation of the nodes only. Since $\Psi(0,\Omega(\lambda))$ is an eigenfunction of $\hat{H}_0$ 
on the nodal domain $\Omega(\lambda)$ we obtain
\begin{equation}
\label{eq5}
\hat{H}_0 \left. \partial_\lambda \Psi(0,\Omega(\lambda)) \right|_{\lambda=0}
= \left.  \partial_\lambda E[0,\Omega(\lambda)] \right|_{\lambda=0} \; \Psi(0,\Omega(0))
 + E[0,\Omega(0)] \; \left. \partial_\lambda \Psi(0,\Omega(\lambda)) \right|_{\lambda=0}
\end{equation}
where the contribution of the second term to the left side of Eq. (~\ref{eq2}) vanishes due to the normalization contraint (~\ref{eq9}).
The modified Hellmann-Feynman theorem for fixed-node quantum Monte Carlo calculations with parameter dependent nodal boundary
is therefore given by
\begin{eqnarray}
\label{eq6}
\left. \partial_\lambda E[\lambda,\Omega(\lambda)] \right|_{\lambda=0} & = &
 \int_{\Omega(0)} \Psi(0) \left. \partial_\lambda \hat{H}_\lambda \right|_{\lambda=0} \Psi(0) \; d\Omega \\ \nonumber
 & - & \frac{1}{2} \int_{\partial \Omega(0)} \mid \! \nabla \Psi(0) \! \mid \left. \partial_\lambda 
 \Psi(0,\Omega(\lambda)) \right|_{\lambda=0} \; d \partial \Omega \\ \nonumber  
 & = & \int_{\Omega(0)} \Psi(0) \left. \partial_\lambda \hat{H}_\lambda \right|_{\lambda=0} \Psi(0) \; d\Omega
+ \left. \partial_\lambda E[0,\Omega(\lambda)] \right|_{\lambda=0} 
\end{eqnarray}
with an additional term which can be interpreted as the linear response of the energy with respect to the variations of the 
the nodal region. This term vanishes \cite{X5} provided that the nodal region $\Omega(0)$ coincides with a nodal region of the 
unconstrained solution of the Schr\"odinger equation. 

Finally we want to mention that the main part of our paper concerning the dipole moment of CO is not affected by 
this correction. In our actual calculations we have only used parameter independent nodal boundaries   
for which the unmodified Hellmann-Feynman theorem remains applicable. However, further studies
of the implications of the additional term seem to be necessary.


\end{document}